\documentclass[prd,preprint]{revtex4}
\usepackage[english]{babel}
\usepackage[utf8x]{inputenc}
\usepackage[T1]{fontenc}
\usepackage{amsmath}
\usepackage{booktabs}
\usepackage{graphicx}
\usepackage[colorinlistoftodos]{todonotes}
\usepackage[colorlinks=true, allcolors=blue]{hyperref}
\title{MESSing with CMB}

\newcommand\SPD{\mathrel{\stackrel{\makebox[0pt]{\mbox{\normalfont\tiny (3)}}}{\Delta}}}

\newcommand\R{\mathcal{R}}

\makeindex

\begin{document}

\title{The MESS of the CMB}

\author{Manuel Alejandro Jaramillo Rodr\'guez${}^{1}$}
\author{Antonio Enea Romano${}^{1,2}$}
\author{Sergio Andr\'es Vallejo-Pe\~na${}^{1}$}

\affiliation{
${}^{1}${ICRANet, Piazza della Repubblica 10, I--65122 Pescara} \\
${}^{2}$Theoretical Physics Department, CERN, CH-1211 Geneva 23, Switzerland
} 

\begin{abstract}
We  analyze cosmic microwave background (CMB) data  taking into account the effects of a momentum dependent effective sound speed (MESS). This approach allows to study the effects of primordial entropy in a model independent way, and its implementation requires a minimal modification of existing CMB fitting numerical codes developed for single scalar field models.

We adopt a phenomenological approach, and study the effects a local variation of the MESS around the scale where other analysis have shown some deviation from an approximately scale invariant curvature perturbation spectrum. We obtain a substantial improvement of the fit with respect to a model without MESS, showing that primordial entropy modeled by MESS can be an explanation of these deviations.
\end{abstract}
\maketitle
\section{Introduction}

According to  the standard cosmological model primordial curvature fluctuations produced during inflation provided the seeds for the CMB (cosmic microwave background) anisotropies and for the formation of large scale structure (LSS). The simplest models of inflation are based on a slowly rolling single scalar field, whose perturbations evolve according to the Sasaki-Mukhanov equation, producing a 
 nearly scale invariant power spectrum of primordial comoving curvature perturbations, which is consistent with CMB observations, but the observational data at same scales also show  some evidence of deviation from scale invariance, which in single scalar field models could be explained for example by temporary violations of the slow-roll conditions \cite{GallegoCadavid:2016wcz,Arroja:2011yu,Romano:2008rr,Chung:2005hn,GallegoThesis,Cadavid:2015iya, 
Gariazzo:2014dla,Motohashi:2015hpa,et,aer,a1,a2,a3,Adams,Chluba:2015bqa,Chen:2011zf,Palma:2014hra, starobinsky,constraints1,
constraints2,Hazra:2014jka,Hazra:2014goa,Martin:2014kja,Romano:2014kla,Ashoorioon:2006wc,Ashoorioon:2008qr,Cai:2015xla}. 

Other possible explanations of these features of the primordial curvature spectrum are  multi-fields models \cite{Romano:2020oov,Braglia:2020fms}, modified gravity theories \cite{Vallejo-Pena:2019hgv}, or a combination of the two. 
A model independent approach to the study of primordial curvature perturbations was recently proposed in \cite{Romano:2018frb}, in which a new equation was derived, generalising the Sasaki-Mukhanov equation to any physical system satisfying Einstein's equation, including multi-fields or modified gravity theories, by introducing two new effective quantities, the momentum dependent effective sound speed (MESS) and the space dependent effective sound speed (SESS).

This approach is  valid for any model in which a total effective stress-energy-momentum tensor (EST) can be defined, including the multi-fields and modified gravity cases, under the provision of moving to the r.h.s of the gravitational field equations the geometrical terms associated to the modification of gravity. One convenient aspect of this approach is the similarity of the equations with the Sasaki-Mukhanov equation, which allows minimal modifications of existing CMB Boltzmann codes, such as CLASS or CAMB, in order to analyse CMB data in a model independent way. In this paper we will show the results of analyzing CMB data with a modified version of CLASS. 

\section{MESS approach}
\label{sec:mess}

For scalar perturbations with respect to a flat  FLRW background, according to the scalar-vector-tensor decomposition (SVT), the most general metric and effective stress energy momentum tensor (EST) perturbations take the form
\begin{align}
ds^2 &= -(1+2A)dt^2+2a\partial_iB dx^idt + a^2\left\{\delta_{ij}(1+2C)+2\partial_i\partial_jE\right\}dx^idx^j \, , \label{pmetric} \\
T^0{}_0 &= - (\rho+\delta\rho) \quad \,, \quad  T^0{}_i = (\rho+P) \partial_i(v+B) \,, \nonumber \\ T^i{}_j &= (P+\delta P)\delta^i{}_j + \delta^{ik}
\partial_{k}\partial_{j}\Pi
-\frac{1}{3} \delta^{i}{}_{j} \SPD \Pi  \, . \label{psem}
\end{align}
where $v$ is the velocity potential and $\SPD \equiv \delta^{kl}\partial_k\partial_l$.

For any physical system admitting a Lagrangian formulation we can take the variation with respect to the metric to obtain the corresponding EST.  As a consequence  all the results derived using eqs.(\ref{pmetric}-\ref{psem}) are general and can be applied to any theory for which an EST can be computed, including multi-fields and modified gravity. 

Note that we have not chosen any gauge in eqs.(\ref{pmetric}-\ref{psem}). The comoving slices gauge, or simply \textit{comoving gauge} for brevity, is defined by the condition $(T^{0}{}_{i})_c=0$, where we denote with a subscript $c$ any quantity evaluated in the comoving gauge. The metric and the perturbed EST in the comoving gauge are denoted as 
\begin{align}
ds^2 &= -(1+2\gamma)dt^2+2a\partial_i\mu\, dx^idt + a^2\left\{\delta_{ij} (1+2\R)+2\partial_i\partial_j\nu\right\}dx^idx^j \, , \label{pmetricC} \\
(T^0{}_0)_c &= - (\rho+\beta) \, , \\ (T^i{}_j)_c&=(P+\alpha)\delta^i{}_j + \delta^{ik}
\partial_{k}\partial_{j}\Pi
-\frac{1}{3} \delta^{i}{}_{j} \SPD \Pi \,.
\end{align}
where we have defined the gauge invariant quantities $\alpha=\delta P_c,\beta=\delta \rho_c,\gamma=A_c,\mu=B_c,\zeta=C_c,\nu=E_c$.

For a single scalar field minimally coupled to gravity, 
the \textit{comoving gauge} and the \textit{uniform field gauge} (or \textit{unitary gauge}) coincide, but they are \textit{different} for more complex  systems. 

In order to the study the evolution of cosmological perturbations in the standard approach \cite{Naruko:2018fwo} entropy perturbations $\Gamma$ are defined according to \cite{Kodama:1985bj} 
\begin{align}
\alpha(t,x^i) &= c_s(t)^2 \beta(t,x^i) + \Gamma(t,x^i) \, , \label{entropy} \end{align}
where $c_s$ is interpreted as sound speed and is a function of time only.
It was shown \cite{Romano:2018frb} that there is an alternative approach which does not require the notion of entropy perturbations, and involves the solution of a single differential equation

\begin{align}
\ddot{\R} + \frac{\partial_t(Z^2)}{Z^2} \dot{\R} &- \frac{v_s^2}{a^2} \SPD \R + \frac{v_s^2}{\epsilon}\SPD \Pi +  \frac{1}{3 Z^2}\partial_{t}\left( \frac{Z^2}{H \epsilon} \SPD \Pi \right)= 0 \, , \label{RcttPi}
\end{align}

where $Z^2\equiv\epsilon a^3/v_s^2$ and an effective space dependent sound speed (SESS) has been defined as

\begin{equation}
v_s^2 (t,x^i) \equiv \frac{\alpha (t,x^i)}{\beta (t,x^i)} \, .\label{vs}
\end{equation}

In this approach the effects of entropy perturbations on curvature perturbations are encoded in the SESS. In fact, we can combine eq.(\ref{entropy}) and (\ref{vs})
to get the relation between entropy and the SESS 
\begin{align}
v_s^2 &= c_s^2\left[ 1 + \frac{\Gamma}{2 H \epsilon \left(\dot{\R} + \frac{1}{3H\epsilon} \SPD \Pi\right)}\right]^{-1} \, . \label{vcgamma}
\end{align}
The definition of entropy perturbations given in eq.(\ref{entropy}) is invariant under the transformations
\begin{align}
c_s^2 & \to \tilde{c}_s(t)^2=c_s(t)^2+\Delta c_s(t)^2 \, , \label{cst} \\ 
\Gamma & \to \tilde{\Gamma}(t,x^i) = \Gamma(t,x^i) - \Delta c_s(t)^2 \beta (t,x^i) \, , \label{Gammat}
\end{align}
where $\Delta c_s(t)^2$ is an arbitrary function of time only. Thus, the definition of entropy is not unique as it is shown by the invariance of eq.(\ref{entropy}) under these transformations.
This \textit{ambiguity} in the definition of the entropy perturbations and sound speed $c_s(t)$ also motivates the introduction of the SESS, which is a \textit{uniquely} gauge invariant quantity.

The momentum dependent effective sound speed (MESS) $\tilde{v}_k(t)^2$ is defined according to
\begin{equation}
\tilde{v}_k^2(t) \equiv \frac{\alpha_k(t)}{\beta_k(t)} \, , \label{ck}
\end{equation}
in order to derive a similar equation to eq.~(\ref{RcttPi}) in momentum space \cite{prep}
\begin{align}
\ddot{\R}_k + \frac{\partial_t(\tilde{Z}_k^2)}{\tilde{Z}_k^2} \dot{\R}_k & + \frac{\tilde{v}_k^2}{a^2} k^2 \R_k - \frac{\tilde{v}_k^2}{\epsilon}k^2 \Pi_k -  \frac{1}{3 \tilde{Z}_k^2}\partial_{t}\left( \frac{\tilde{Z}_k^2}{H \epsilon} k^2 \Pi_k \right)= 0  \, , \label{Rckeq}  
\end{align}
where $ \tilde{Z}_k^2\equiv\epsilon a^3/\tilde{v}_k^2$.
Since the product of the Fourier transforms of two functions is the transform of the convolution of the two functions, the MESS $\tilde{v}_k(t)$ defined in eq.(\ref{ck}) is not the Fourier transform of the SESS $v_s(x^{\mu})$ defined in eq.~(\ref{vs}).

In this paper we will consider the case of an isotropic EST,
for which eq.~(\ref{Rckeq}) reduces to
\begin{align}
\ddot{\R}_k + \frac{\partial_t(\tilde{Z}_k^2)}{\tilde{Z}_k^2} \dot{\R}_k & + \frac{\tilde{v}_k^2}{a^2} k^2 \R_k = 0  \, . \label{Rckeq2}  
\end{align}
It can be shown that the Sasaki-Mukhanov equation can be obtained from eq.~(\ref{Rckeq2})
when $\tilde{v}_k$ is a function of \textit{time only}.

It is important to note that the MESS approach can be applied to any system for which an appropriate EST can be obtained, including multiple fields, modified gravity theories, combinations of the two, and supergravity, and is particularly useful for model independent analysis in which the MESS is treated as a phenomenological quantity constrained by the data.

\section{Inflationary model}
In order to solve the perturbations equation we need to define the evolution of the background. For definiteness we model the background with a single minimally couple scalar field, but we solve the MESS perturbation equation shown in the previous section.
The action for a scalar  is given  by
\begin{equation}
    S = \int dx^{4}\sqrt{-g}\left(\frac{R}{16\pi G} + \frac{1}{2}g^{\mu \nu }\partial_{\mu }\phi \partial_{\nu } \phi - V(\phi)\right)
\end{equation}
where $R$ is the Ricci scalar, $g^{\mu \nu}$ is the FLRW metric and $V(\phi)$ is the inflation potential. By taking the variation with respect to the metric we can obtain the Friedmann's equation
\begin{equation}
    H^{2} = \left(\frac{a}{\dot a}^{2}\right) = \frac{8\pi G}{3}\left(\frac{1}{2}\dot \phi^{2} + V(\phi)\right) \,,
\end{equation}
and the variation with respect to the field gives 
\begin{equation}
    \ddot \phi + 3H\dot \phi + \frac{\partial V(\phi)}{\partial \phi} = 0 \,.
\end{equation}
The slow-roll parameters are defined as
\begin{equation}
\begin{aligned}
    \epsilon \equiv -\frac{\dot H}{H^{2}} \\
    \eta \equiv \frac{\dot \epsilon }{\epsilon H}
\end{aligned}
\end{equation}
In this paper we will consider background models with a quadratic potential
\begin{equation}
    V(\phi) = V_{2}\phi^{2}  = \frac{1}{2} m^{2} \phi^{2} \, ,
\end{equation}
where $m$ is the inflaton mass.

There is no fundamental reason behind the choice of this potential, we just use it as an example for applying the MESS approach, and show that a local variation of the effective sound speed leads to a substantial improvement of the data fitting. 
\section{Data analysis using the MESS approach}
It's known from the  Planck mission \cite{Akrami:2018odb} data analysis  that there are some features in the primordial scalar spectrum at $k \approx 0.002 Mpc^{-1}$ and $k \approx 0.0035 Mpc^{-1}$ \cite{Akrami:2018odb}, and in this paper we will show how they can be explained by a local variation of the MESS. 
In order to study the effects of the MESS on the primordial spectrum we modified the Boltzmann solver code CLASS \cite{brinckmann2018montepython} and introduced a  local momentum dependency in the MESS around the scale where the main features appear in the primordial spectrum. The fit with  different inflationary models and a more general parametrization of  the MESS is left for future works.
We use data sets from the Planck collaboration (Planck 2018 data) \cite{Akrami:2018vks}, and since the MESS is affecting scalar perturbations we focus our attention on the temperature spectrum.

We modify the Primordial module of CLASS, adding a new routine named "$primordial\_inflation\_v\_k$" that computes the MESS and uses it to solve eq.(\ref{Rckeq}) for  curvature perturbations in absence of anisotropies, for the case when the effective sound speed is not time dependent, which takes the form 
\begin{align}
\R_k''+\frac{\partial_{\eta}(z^2)}{z^2}\R_k'+\tilde{v}_k^2k^2\R_k &=0 \, , \label{Rcpk} \\
u_k''+\left(\tilde{v}_k^2 k^2 -\frac{z''}{z} \right)u_k&=0 \, , \label{ukp}
\end{align}
where the prime denotes derivatives with respect to conformal time $\tau$, and $z^2=2a^2\epsilon$.

The initial conditions for curvature perturbations in presence of MESS are taken according to \cite{Romano:2018frb} 

\begin{equation}
\begin{aligned}
    u_{\kappa}(\tau_{0}) = \frac{1}{\sqrt{2\kappa \tilde{v}_k}} \, ; \,
    u'_{\kappa}(\tau_{0}) = - \frac{\tilde{v}_k k}{\sqrt{2\kappa}} \,.
\end{aligned} 
\end{equation}

\begin{figure}[htb!]
\centering
\includegraphics[width=1\textwidth]{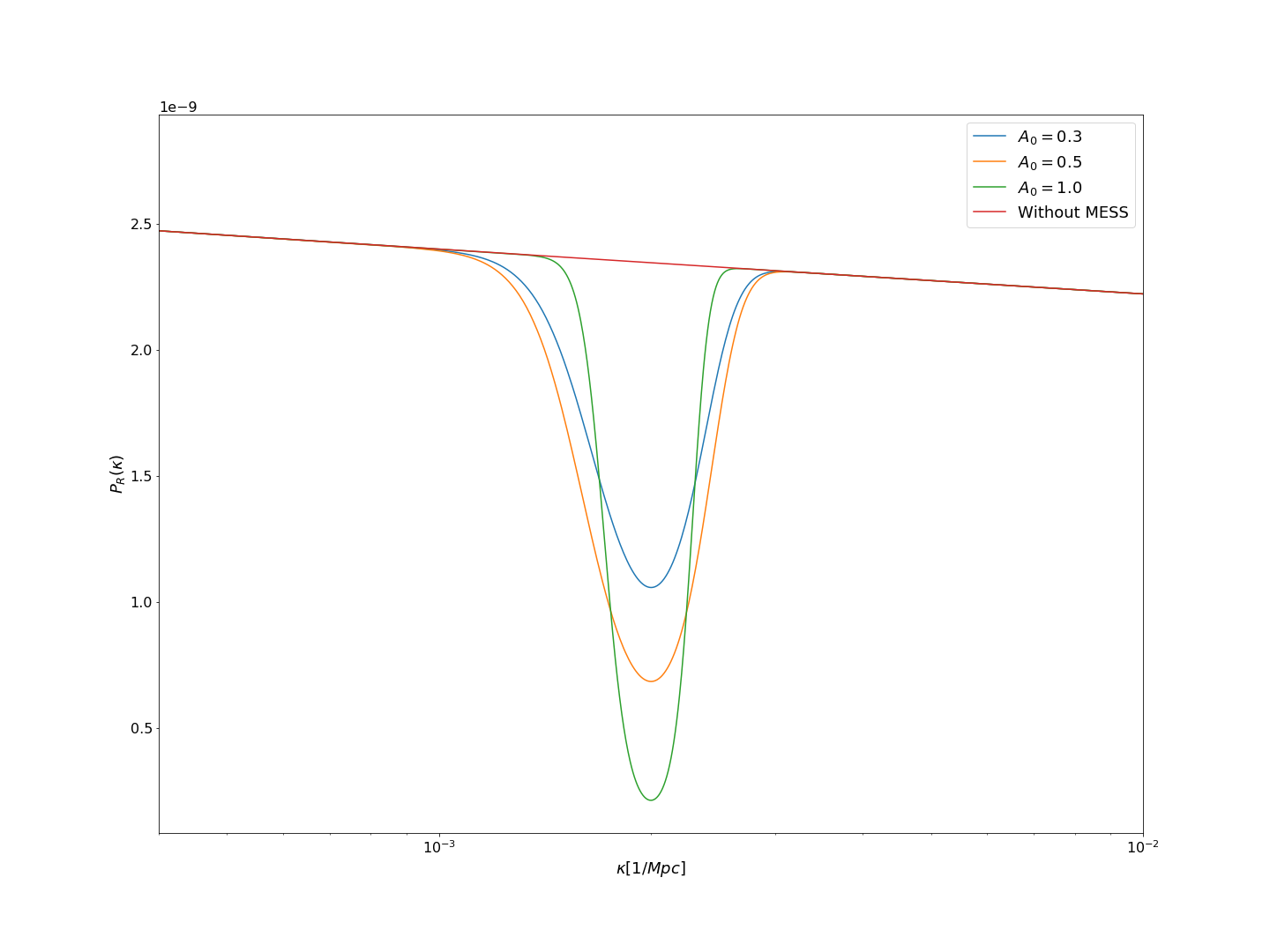}
\caption{\label{fig:amplitud}The effects on the power spectrum of the MESS, as parametrized in eq.(\ref{vkloc}), are shown  for different values of $A_{0}$, with  $k_{0} = 2 \times 10^{-4} Mpc^{-1} $ and $\sigma_{0}= 3 \times 10^{-5}$.}
\end{figure} 

\begin{figure}[htb!]
\centering
\includegraphics[width=1\textwidth]{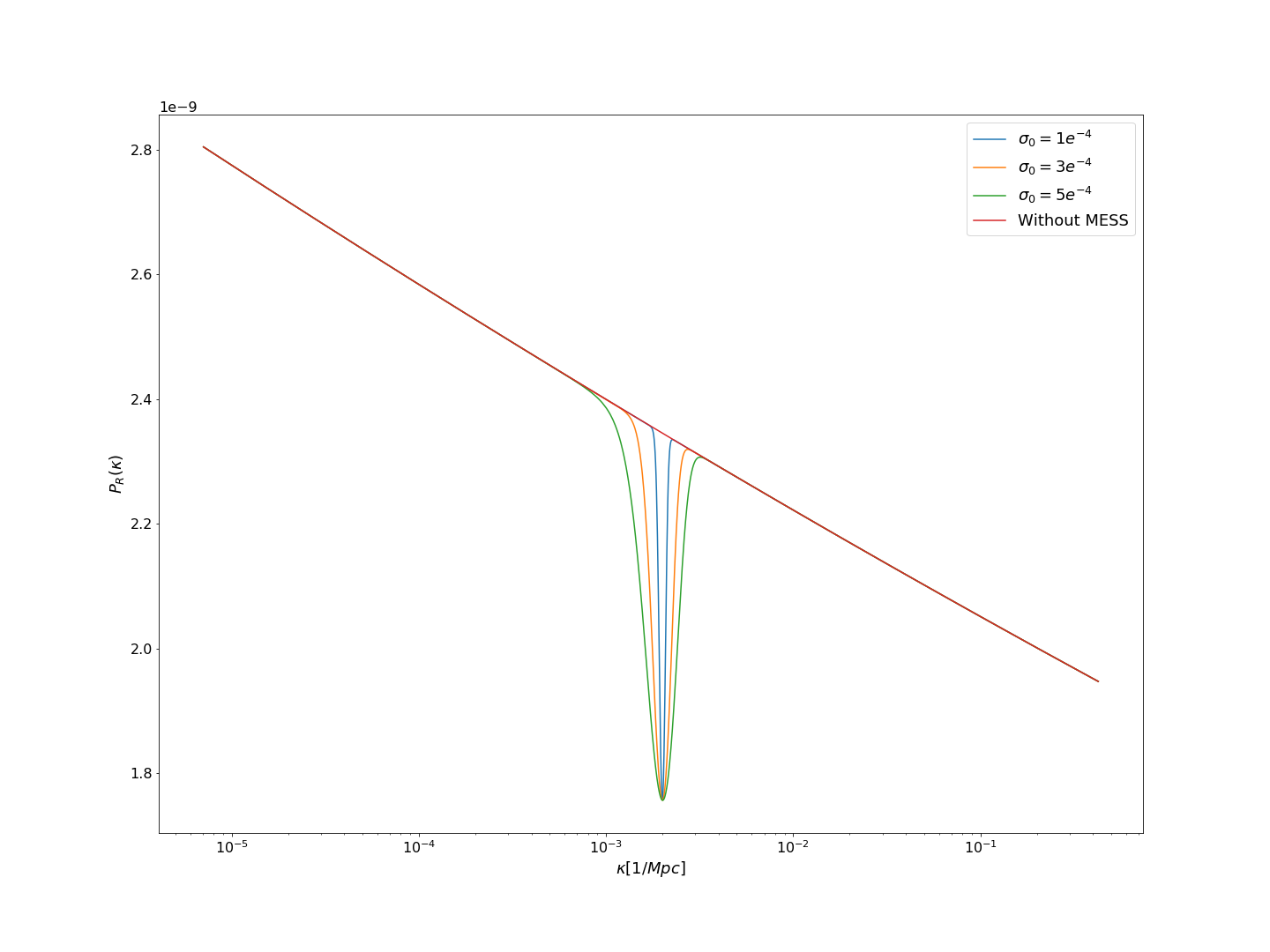}
\caption{\label{fig:sigma}The effects on the power spectrum of the MESS, as parametrized in eq.(\ref{vkloc}), are shown  for different values of $\sigma_{0}$, with  $k_0 = 2 \times 10^{-4}  Mpc^{-1} $ and $A_{0}= 0.1$.}
\end{figure}

\begin{figure}[htb!]
\centering
\includegraphics[width=1\textwidth]{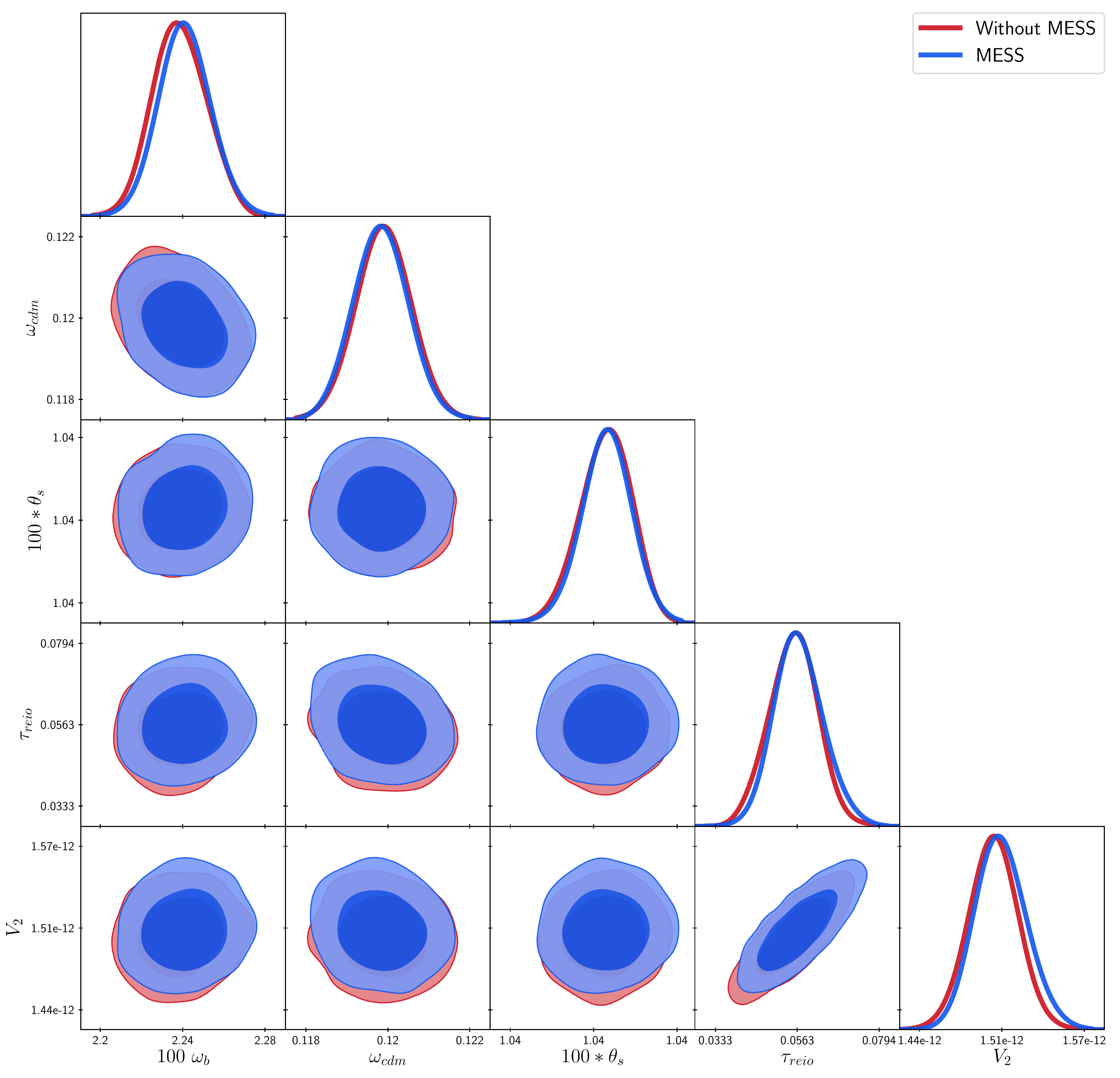}
\caption{\label{fig:cont}Comparison of the background parameters for a model with and without MESS. As can be seen the effect of MESS is negligible.}
\end{figure}

\begin{figure}[htb!]
\centering
\includegraphics[width=1\textwidth]{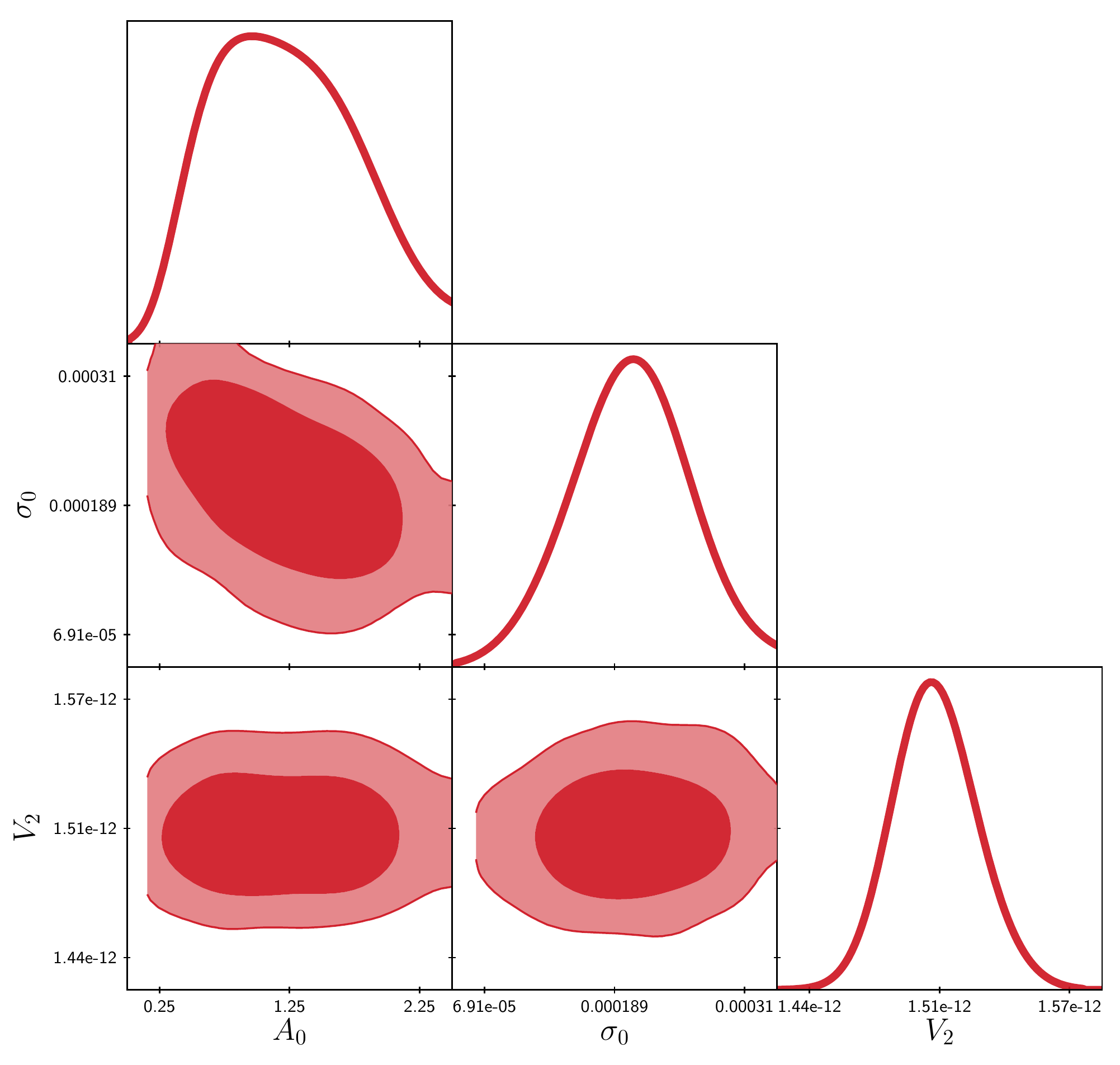}
\caption{\label{fig:feature}Results of the data fitting analysis for the MESS parameters $A_0$ and $\sigma_0$.}
\end{figure}

\section{Relation between primordial and late time entropy}
It is important to observe that the entropy  producing the momentum dependency of the MESS is primordial, i.e. it is related to physical phenomena taking place during inflation, before reheating. This  entropy is not directly related to the entropy of the CMB plasma, and is only imprinted in the primordial power spectrum which is used as input in the calculation of the CMB spectrum. The CMB codes such as CLASS are also solving equations involving the entropy of the CMB plasma, related to the different species composing it, but since the reheating process can be very complicated, it is not easy to relate the primordial to the late time entropy.  Consequently primordial entropy can be investigated primarily from its effects on the primordial curvature spectrum, which can be studied in a model independent way using the MESS, independently from the constraints on the late time entropy of the plasma. 

The conversion of primordial into late time entropy during reheating goes beyond the scope of this paper, but due to its generality, the MESS approach could also be used to study the evolution of curvature perturbations during reheating.
\begin{figure}[htb!]
\centering
\includegraphics[width=1\textwidth]{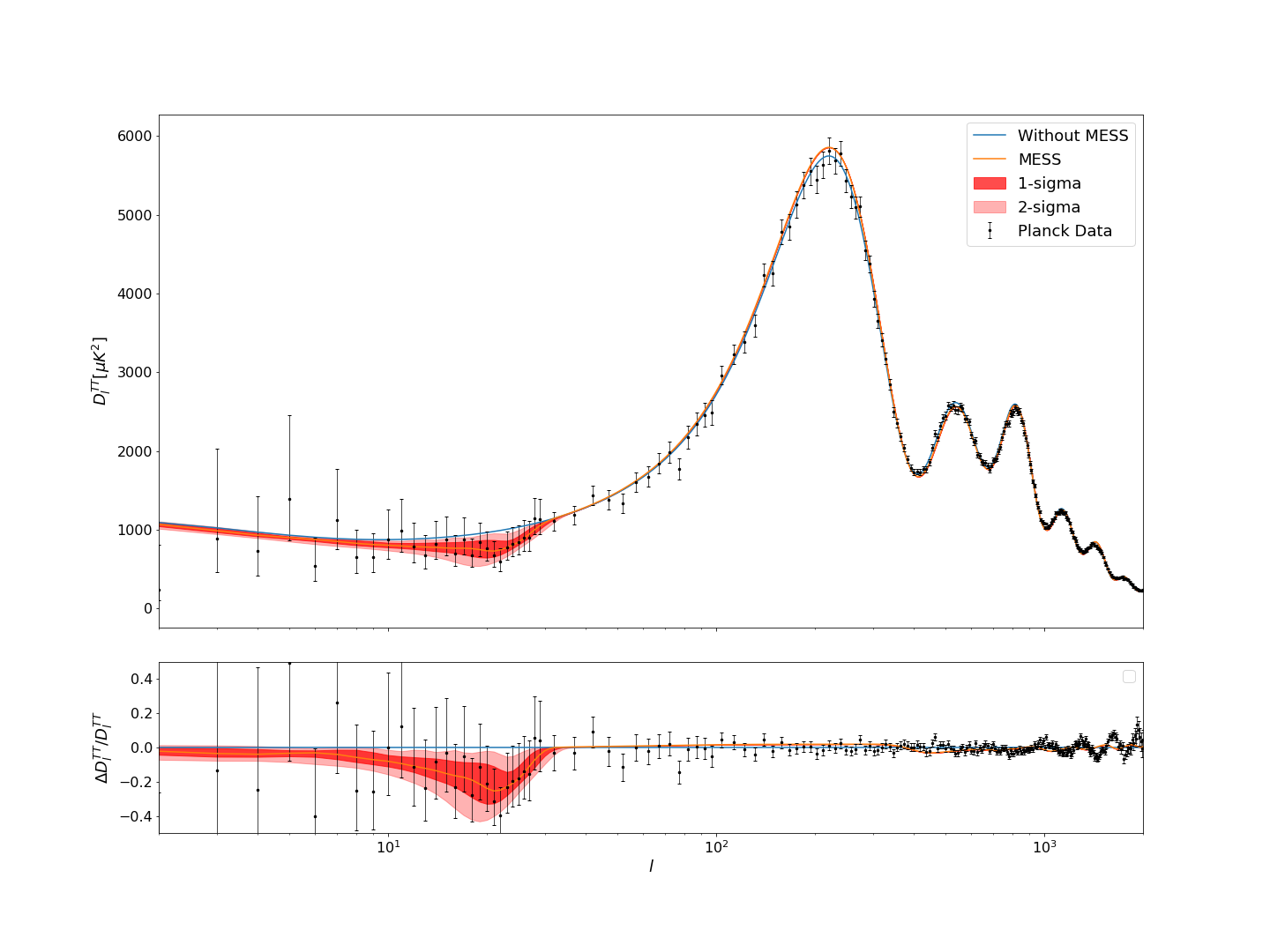}
\caption{\label{fig:TT}In the upper panel it is plotted the CMB temperature spectrum $D_l^{TT}=\ell(\ell+1) C_\ell^{TT}/(2\pi)$, and in the lower panel  the relative difference between a model with and without MESS. As can be seen the model with MESS is fitting the data better in the range $10<l<30$.  }
\end{figure}

\section{MESS parametrization}
We fit the data with a local parametrization of the MESS given by
\begin{equation}
\tilde{v}_k = 1 + A_{0}e^{-(\frac{\kappa - \kappa _{0}}{\sigma _{0}})^{2}} \label{vkloc}
\end{equation}
where $\kappa _{0}$ is the scale of the feature, $\sigma_{0}$ and $A_{0}$ are the standard deviation and the amplitude of the Gaussian used to model the MESS. When $A_{0} = 0$ we recover the standard value of the sound speed $\tilde{v}_k = 1$ .
For the background  cosmological parameters, i.e. the current density of baryons $\Omega _{b}$, the current density of Cold Dark Matter $\Omega _{cdm}$, the reionization optical depth $\tau$, the acoustic angular scale $\theta$ and the coefficient of the quadratic term of the potential $V_{2} = \frac{1}{2} m_{i}^{2} $, we use priors given by the best fit values obtained in Planck data Analysis  \cite{Aghanim:2018eyx}.
We will focus on the main feature of the primordial power spectrum, that is a dip located at $k \approx 0.002 Mpc^{-1} $, so we fix $k_0$ to that value.

For $\tilde{v}_k > 1$ we obtain a dip of the primordial spectrum, and for $\tilde{v}_k < 1$ a bump, as shown in Fig(\ref{fig:amplitud}), where the spectra corresponding to different values of $A_0$ are plotted. In Fig(\ref{fig:sigma}) are shown the effects of different values of $\sigma _{0}$ on the primordial power spectrum. As can be seen the width of the MESS $\sigma _{0}$ is related to the width of effect on the power spectrum, while the magnitude of the MESS $A_0$ is related to the amplitude of the effect on the power spectrum.

\section{Results of the data analysis}
In table(\ref{tab:res}) we show the best-fit parameters with and without MESS. The fit with MESS is producing an improvement of the reduced  $\chi^{2}$ of about 12 with respect to the model without it. 
In Fig.(\ref{fig:cont}) we show the marginalised contour plots for the background cosmological parameters for the models with and without MESS. As can be seen the effect of the MESS on the estimation of these parameters is negligible. Fig(\ref{fig:feature})  shows that $\sigma_{0}$ and $A_{0}$ are approximately inversely proportional, i.e. large values of $A_{0}$ are associated to  a smaller  width of the Gaussian $\sigma_{0}$.

\begin{table}[htb!]
\centering
\begin{tabular}{@{}lll@{}}
Parameter         & Without MESS         & MESS             \\ \midrule
$100\Omega_{b}$   & $2.236 \pm 0.012$   & $2.238 \pm  0.012$  \\
$\Omega_{CDM}$    & $0.119 \pm 0.001$ & $0.119 \pm 0.001$ \\
$\tau$            & $0.056 \pm 0.006$  & $0.057 \pm 0.006$ \\
$\theta$          & $1.0403 \pm 0.0002$ & $1.0340 \pm 0.0002$ \\
$10^{12}V_{2}$ & $1.5003 \pm 0.0196$ & $1.5069 \pm 0.0206$ \\ \midrule
$10^{5}\sigma_{0}$     & ---                 & $2.08^{+0.59}_{-0.61}$ \\
$A_{0}$         & ---                 & $1.29^{+0.50}_{-0.73} $                  \\ \midrule
$\chi^{2}$        & $2789.58$           & $2777.42$    
\end{tabular}
\caption{\label{tab:res}The results of the data fitting with and without MESS are shown for the background and MESS parameters, as defined in eq.(\ref{vkloc}). The reduced $\chi ^{2}$ of the model with MESS is improved by about 12 with respect to the model without MESS.}
\end{table}
Fig(\ref{fig:TT}) shows the effect of the MESS on the CMB temperature  spectrum. From the residuals plot we can see that the MESS is improving significantly the fit around the $l=20$, with an improvement of the reduced $\chi^2$ of about $12$.
\section{Conclusions}
We have analyzed CMB data with a modified version of the CLASS code, taking into account the effects of a momentum dependent effective sound speed. This approach allows to study the effects of primordial  entropy in a model independent way, and its implementation requires a minimal modification of existing codes developed for single scalar field models.

We have considered a local variation of the MESS around the scale where other analysis have shown some deviation from a scalar invariant curvature perturbation spectrum, obtaining a substantial improvement of the fit.

In the future it will be interesting to extend the analysis to other more general parametrizations of the MESS, such as the  Piecewise Cubic Hermite Interpolating Polynomial (PCHIP) to investigate systematically possible variations of the MESS at any scale.
It could also be interesting to consider the case when the MESS is  time dependent, or other background models. 

\acknowledgments
We thank Stefano Gariazzo and Krzysztof Turzyński for interesting discussions.
\bibliographystyle{h-physrev4.bst}
\bibliography{sample}
\end{document}